# Multi-Utility Market: Framework for a Blockchain Exchange Platform for Sustainable Development

Jacques Bou Abdo[1], Sherali Zeadally[2]

[1] Notre Dame University, Lebanon, jbouabdo@ndu.edu.lb

[2] University of Kentucky, USA, szeadally@uky.edu

Water and other resources are becoming scarcer every day, and developing countries are the neediest for an immediate intervention. Water, as a national need, is considered to be one of the main causes for conflicts in the 21st century. Peer-to-peer trading is one of the most convenient, scalable and sustainable solutions but faces organization challenges such as: the absence of suitable business models motivating normal users to sell their generated resources, currency and financial settlement complexities, and single utility markets. We propose a multi-utility trading platform, based on blockchain technology which can address the challenges faced by peer-to-peer trading. This platform meets the needs of developing countries in particular as well as rural areas of developed countries. The open nature of our proposed design makes it suitable for adoption and use by various stakeholders.

## Introduction

Water supplies are becoming scarcer every day. Developing countries are not able to buy additional supplies from neighboring countries. 44 countries have high to extremely high risk of water crisis [1] [2] out of which, 28 are considered developing countries by the United Nations [3] which has listed its urgent sustainability goals and energy resource management among its most pressing goals [4].

On 25 September 2015, the general assembly of the United Nations approved the resolution: “Transforming our world: the 2030 Agenda for Sustainable Development” [4] which set 17 goals and 230 indicators as global objectives expected to guide the actions of the international community towards global sustainability. Out of these 17 goals, 4 are related to resource management and its direct consequence. The energy-related goals are:

- Goal 2: End hunger, achieve food security and improved nutrition and promote sustainable agriculture.
- Goal 6: Ensure availability and sustainable management of water and sanitation for all.
- Goal 15: Protect, restore and promote sustainable use of terrestrial ecosystems, sustainably manage forests, combat desertification, and halt and reverse land degradation and halt biodiversity loss.

ICT solutions for water management has become extremely important as a global initiative and this motivates our work [49]. Scientists and policy makers are questioning the sustainability of conventional urban water supply systems where drinking water is pumped from underground sources and wastewater treatment occurs at distant locations [5]. Governments in developing countries are rarely capable of developing and maintaining nationwide resource management infrastructures. As a result, they often develop a private, distributed, and delegated infrastructure as the most convenient solution especially for rural areas [48] [49]. New demand-driven water delivery models enable urban consumers to also become prosumers within the system, by integrating hybrid water systems into their water consumption mix and trading excess water with their peers, either physically or via a ‘credit-debit’ scheme [5]. The guiding principles for such new models are [5]:

- Utilize previously unused water sources collected locally.

- Encourage the use of alternative water technologies such as rainwater collection systems.
- Improve the management of existing shallow groundwater resources.
- Minimize the operating costs of existing water systems.
- Invest in new large infrastructure developments.
- Maintain water’s nutrient resources after water management systems.
- Create a marketplace for trading, in the form of physical peer-to-peer water trading and/or via a newly developed credit-debit system.

In [52], Stewart et al. proposed a smart multi-utility service provider that relies on intelligent measurements and decisions to deliver its services.

Although the water crisis is reaching epic proportions, water exchange markets have been barely discussed in literature. Existing solutions are still in the infancy stages, especially when the business models motivating parties to adopt these solutions have hardly been explored. Energy trading platforms have been heavily discussed [31] [33] [34] [35] [36], but having a per-utility platform is not the optimal solution. To address this issue, in this work, we propose a multi-utility market framework for exchanging resources/utilities. To the best of our knowledge, this is the first utility trading platform that takes into consideration the specific needs of developing countries and the first multi-utility trading framework to be proposed in the literature. Our proposed framework is based on the lessons learnt from the current literature. This framework attempts to address many of the problems associated with financial settlements, transparent pricing, smart metering, multi-resources, smart contracts, delegation and integration. Additionally, the proposed framework should require minimum maintenance from municipalities and other authorities and it depends on private initiatives for sustainability. As seen above, the majority of the countries facing high risk of water crisis are developing, and those are most desperate due to the absence of public strategies. A platform not catered towards the needs of developing countries may suit developed one, but leave the ones most needing it. A successful platform should allow business model innovations from the private sector. Local markets should get the highest priority due to their lower costs in transporting energy and water [6]. A successful platform should be able to solve financial settlements in a simple way.

***Contributions of this work***

We summarize the main contributions of this paper as follows:

(1) This is the first multi-utility framework proposed in literature.
(2) It responds to the requirements of rural communities and developing countries, which only requires minimal investment and maintenance from the state and attractive business models for private initiatives.
(3) It solves the weaknesses identified in existing utility frameworks.

The remainder of this paper is organized as follows. Section 2 surveys the literature and identifies important features in a successful resource management system and existing exchange frameworks. Section 3 specifies the requirements needed by a successful utility trading system and proposes our framework that satisfies the requirements identified. Section 4 discusses some of the challenges that need to be addressed when implementing this framework. Finally, we make some concluding remarks in section 5.

## Related Work

### *Water Trading*

Grigoras et al. [7] proposed a blockchain-based framework for dynamic water pricing options and policy actions as a function of the energy prices. In other words, the water prices drop as the price of the energy needed to pump the water drops. Smart measurement and dynamic energy pricing help the platform achieve significant energy savings, and thus supply the consumers with quality water at an affordable price. The main contribution of this paper is related to pricing rather than trading.

Chohan [8] presented the IBM's water management system as a case study of Blockchain's environmental sustainability's role. Pee et al. [9] proposed a lightweight peer-to-peer water trading system facilitating transactions between irrigators and consumers. The proposed system uses private Ethereum blockchain and Geth (**G**o-**eth**ereum) command line interface [45], which is a Go language implementation of Ethereum. It is not clear whether the proposed system uses Ethereum's native currency 'Ether' or a higher level fiat currency.

Dogo et al. [10] studied the impact of using blockchain and IoT technologies on intelligent water management systems from an African perspective. This work considers the requirements (continuous monitoring, maintenance prioritizing, remote control of water distribution and transparent and confidential compliance with regulatory and policy requirements) African countries generally need to satisfy with the deployment of new water management systems. This work will help us select the requirements for comparing water management systems as shown in section 2.3. In the same context, Lin et al. [12] proposed an evaluation tool to determine the context-specific requirements (technical and social) for blockchain-based ICT e-agriculture systems. In other words, the authors proposed a flowchart that helps ICT e-agriculture system designers select a suitable blockchain type.

Poberezhna [13] showed the advantages of green economics and Blockchain and their possible integration and studied water credits and water trading on Blockchain as examples. Patil et al. [14] proposed a lightweight blockchain based architecture to be used in smart greenhouse farms. This architecture is not concerned with trading, but focuses on the design of a secure platform that can be used as a building block in larger blockchain-based trading applications. In the same context, Ntuli and Abu-Mahfouz [25] proposed a security architecture for smart water management systems. Similarly, the proposed system is expected to be used as a building block in other water management or trading systems.

Lou et al. [24] proposed an Inexact Two-stage Stochastic Nonlinear Programming (ITSNP) model for water resources management and trading under the uncertain conditions (uncertain water supply and water demand). This model can help smart water management or trading systems set the price, based on future consumption and availability predictions. In the same context, Grizzetti et al. [15] and Poff et al. [16] proposed ecological performance metrics (such as floodplain performance threshold and flow recession rate) concerned with an intelligent water management system's efficiency in dealing with unknown future hydrological and climate states.

Antonucci et al. [11] surveyed blockchain applications in the agrifood sector and concluded that these applications are very promising and have great potential but are still immature and hard to apply due to the complexity of blockchain applications. It is not clear how the authors reach this conclusion on complexity.

Several use cases of water management have been studied for Africa [17] [18], Brazil, Mexico, Thailand, the United States [19] [20] and the world [21]. Biswas [22] showed that Integrated Water Resources Management (IWRM) is exceedingly difficult to be implemented in real life, especially that monitoring, management and maintenance operations, on national scale, required for seamless operation are more expensive than legacy water distribution systems. This motivated us to propose a delegated system where most of the generation and consumption sites are privately owned and maintained. Jacobs et al. [23] evaluated and compared the role of participatory governance and scientific information in decision-making in four basins in Brazil, Mexico, Thailand, and the United States. They observed that end-user participation works better in short-term decisions, such as water allocation, compared to long-term decisions such as infrastructure implementation. They also observed that an important reason for IWRM implementation failure is the lack of wide recognition of the expensive capacity building process which allows meaningful stakeholder engagement in water-management decision processes.

***Energy Trading***

Wu and Tran [26] surveyed the applications of blockchain technology in sustainable energy systems and listed the following applications:

- *P2P energy transaction*: The blockchain-based P2P energy trading model can provide an efficient, inexpensive, open, and trustworthy trading platform for the Energy Internet.
- *Electric vehicle*: Incompatible electric vehicles charging pile and payment platforms is an obstacle facing the widespread emergence of green transportation. Blockchain has the capacity to integrate between different platforms thus solving the compatibility issue and alleviate the status of a small number of charging piles using on-time lease of private charging piles based on smart contracts and distributed general ledger technology.

- *Physical information security*: Blockchain's immutability, decentralization, high-redundancy storage, high security, and privacy protection can be a corner stone for solving the security problems faced by information and physical systems.
- *Carbon emissions certification and trading*: Blockchain can facilitate carbon dioxide emission rights exchange between companies.
- *Virtual power plant*: Blockchain plays an important role in the aggregation of distributed generation resources and the establishment of virtual power resource transactions.
- *The synergy of the multi-energy system*: Blockchain can be used to integrate multi-energy system either by developing a decentralized system [27] a complimentary system [28] or a collaborative autonomous system [29].
- *Demand side response*: Blockchain can solve the problem of false accounting and wrong accounting in automatic demand response services, it can also establish a complete set of traceability systems and supervise the settlement of each participating transaction fund.

Basden and Cottrell [30] presented the applicability of blockchain in the management of distributed energy resources. The blockchain applications that have been proposed are as follows:

- *Industry flows*: settlements, change of supplier and real-time capacity matching.
- *Asset management*: Autonomous network configuration, Self-serve maintenance, Asset and inventory tracking and cross asset/industry data sharing.
- *Identity management*: Eligibility for social tariffs, Safety authorizations and permit to work and Fraud detection.

- *Smart contracts*: Electric vehicle charging, Peer-to-peer trading and Demand side management.

Mengelkamp at al. [31] presented a market design and simulation of a local energy market between 100 residential households. The proposed design uses a private blockchain, which underlines the decentralized nature of local energy markets. This design provides energy prosumers and consumers with a decentralized market platform for trading local energy generated without the need for a central intermediary. Similarly, Mattila et al. [32] developed a tentative use case for autonomous machine-to-machine electricity transactions in a housing society environment. Such use cases, are the building blocks for the business model of any energy market.

Di Silvestre et al. [33] studied the power losses in microgrids which could increase with the use of blockchain-based energy markets. The same authors depicted in [34] a comprehensive framework for enhanced microgrid management in the blockchain era and considered for the first time, the provision of ancillary services. The additional costs resulting from ancillary services are analogic to dynamic mining fees. In other words, the power losses resulting from energy exchange is charged, as ancillary services or dynamic mining fees, on top of the respective transactions. In an another work, Di Silvestre et al. [35] studied the possibility to use a smart contract for defining a distributed Demand-Response mechanism. Such a system allows the consumers to request a specific energy resource ahead of time, and then the producer generates the requested resources only, thus minimizing energy losses. A discount system can be built on top of Demand-Response systems allowing the producer to sell the non-reserved resources with a discounted price.

Energy providers tend to normalize the demanded energy over time by providing price incentives. Pop et al. [36] investigated the use of blockchain mechanisms in normalizing the energy demanded by Distributed Energy Prosumers. They also proposed a consensus-based validation for Demand-Response programs and the activation of appropriate financial settlements. In the same context, Devine and Cuffe [39] proposed a mechanism, for blockchain electricity marketplaces, intended to reward organic price-responsive load shifting by incentivizing the consumption of electricity when it is locally abundant.

Dorri et al. [37] proposed a secure and private blockchain-based framework for energy trading in a smart grid. Price negotiation, privacy, authentication of anonymous smart meters and Trusted Third Party (TTP) independency are provided by the proposed framework. To facilitate the implementation of the proposed framework, the same authors [38] proposed a solution to another fundamental challenge in smart grid, which is the security and privacy in Direct Load Control (DLC). They solved it by proposing a blockchain-based framework which allows participating node share their data with the distribution company in an anonymous and secure manner.

Li et al. [40] proposed a secure peer-to-peer (P2P) energy trading platform for industrial Internet of things (IIoT) based on Consortium Blockchain. They also proposed an optimal pricing strategy using Stackelberg game for credit-based loans. Aitzhan and Svetinovic [41] also proposed a secure and private decentralized energy trading through multi-signatures, blockchain and anonymous messaging streams.

Sikorski et al. [42] proposed a machine-to-machine electricity market based on blockchain for the chemical industry. This proposal was among the first to discuss a fully automated blockchain-based electricity exchange market.

***Comparison of Utility (Energy/Market) Exchange Platforms***

Table 1 compares the surveyed frameworks based on the 10 key requirements, we found across the literature, to be important for a resource management system. The considered requirements will be elaborated in section 5, but will be only introduced next. The 10 considered requirements are:

(1) *Financial Settlement (FS)*: whether the trading platform guarantees the transfer of fund when the service is delivered and guarantees the non-transfer of fund if the service is not delivered.

(2) *Transparent Pricing (TP)*: whether the pricing schemes used are fair and the user's accessibility to all offers is guaranteed.

(3) *Smart Contacts (SC)*: whether smart contracts are enabled. Smart contacts are automatic functionality to agree on the service and payment transaction and to enforce the application of this functionality.

(4) *Smart Metering (SM)*: whether smart metering is applied. Smart metering is per service verified, signed and non-repudiated measurements.

(5) *Delegated (Del)*: whether the supply rights are delegated to all users i.e. whether any consumer is a potential prosumer.

(6) *Multi-resource (MR)*: whether the trading platform accepts multiple utilities.

(7) *Blockchain-based (BC)*: whether transactions are booked using a blockchain.

(8) *Business Model Opportunities for private initiatives (BMO)*: whether the trading platform is open to innovations in the business model.

(9) *Integration between incompatible interfaces (INT)*: whether the trading platform has integration middleware for connecting incompatible devices.

(10) *Security and Privacy (SP)*: whether security and privacy are maintained and considered.

Table 1. Comparison of existing water and energy exchange platforms.

| | Platform | FS | TP | SC | SM | Del | MR | BC | BMO | INT | SP | Strength | Weakness |
|---|---|---|---|---|---|---|---|---|---|---|---|---|---|
| Water exchange platforms | [7] | No | No | No | Yes | No | No | Yes | No | No | No | AI-based water pricing options realizing significant energy savings in water delivery | Only suitable for a centralized infrastructure |
| | [8] | Yes | Yes | Yes | Yes | No | No | Yes | No | No | No | Suitable for a unidirectional market | Unidirectional service: users are only consumers or retailers not generators |
| | [9] | Yes | Yes | Yes | Yes | No | No | Yes | No | No | No | Suitable for a unidirectional market | Unidirectional service: users are only consumers or retailers not generators |
| | [14] | No | No | Yes | Yes | No | No | Yes | No | No | Yes | Focuses on the security and privacy | Did not address many of the considered features |
| Energy exchange | [31] | Yes | Yes | Yes | Yes | Yes | No | Yes | No | No | No | One of the most technically developed frameworks | Vulnerable to currency exchange, supports one |
| | | | | | | | | | | | | | resource type only |
| | [32] | No | Yes | Yes | Yes | No | No | Yes | No | No | Yes | Focuses on the Swedish grid pricing (grid access points and electricity consumption) | Does not consider establishing local energy markets; Users interact with the centralized grid only |
| | [34] | No | No | Yes | Yes | No | No | Yes | No | No | No | First to discuss ancillary services for Energy Blockchain for Microgrids. Most developed in energy distribution | Does not discuss financial settlement, pricing, delegation and other important features; does not discuss the business model or blockchain type |
| | [36] | Yes | Yes | Yes | Yes | Yes | No | Yes | No | No | Yes | Includes Demand Response Programs and Penalties for non-compliant users | Does not propose integration between incompatibl e technologie s; does not include |
| | | | | | | | | | | | | | other forms of resources other than energy |
| | [37] | Yes | No | Yes | Yes | No | No | Yes | No | No | Yes | Doesn’t rely on a Trusted Third Party | Does not discuss price negotiation, service searching and other important features |
| | [39] | No | Yes | Yes | Yes | No | No | Yes | No | No | No | Reward organic price-responsive load shifting | Does not discuss prosumer incentivization |
| | [40] | Yes | Yes | Yes | Yes | Yes | No | Yes | No | No | Yes | Includes credit-based payment | Supports one resource type only; does not address integration between incompatible technologies |
| | [41] | No | Yes | Yes | Yes | Yes | No | No | No | No | Yes | Good power-to-security ratio | Supports one resource type only; does not address integration |
| | | | | | | | | | | | | | between incompatible technologies |
| | [42] | No | Yes | Yes | Yes | Yes | No | Yes | No | No | Yes | Controlled energy generation | Supports one resource type only; does not propose integration between incompatible technologies |
| ***Our proposed framework*** | | *Yes* | *Yes* | *Yes* | *Yes* | *Yes* | *Yes* | *Yes* | *Yes* | *Yes* | *Yes* | *First multi-resource framework. First to consider integration interface for incompatible devices. Proposes a dynamic business model that motivates private parties to invest in the platform thereby reducing reliance on* | *It does not include any price prediction algorithms* |

| | | | | | | | | | | | | |
|---|---|---|---|---|---|---|---|---|---|---|---|---|
| | | | | | | | | | | | *public initiatives* | |

### *Gap Analysis*

As table 1 shows, none of the available trading platforms meet all the requirements and each platform has more than one weakness. This makes them impractical, even though many are technically sound, but their business models have not been developed and such inconvenience will likely cause the platform to fail in practice. Existing platforms focus mainly on security, transparent pricing and financial settlement but neglected other important requirements. We found three major weaknesses in every platform in the recent literature. The first weakness is that they only support one type of resources or utility. A water trading platform supports only water trading, and energy trading platform only supports energy trading. This causes multiple incompatible systems to interoperate. The second weakness is the lack of openness [51] in the proposed platforms. Proprietary or closed platforms are usually very efficient in terms of performance but limit the possibilities for new services and wider adoption. The third weakness is also due to lack of openness in the recently proposed platforms. None of these platforms included an integration interface for incompatible devices to be connected. This problem has been reported in the literature [26], but none of the proposed platforms have addressed this weakness.

Another common weakness is delegation, where the majority of the surveyed platforms (8 of 13) differentiate between suppliers and consumers without having the possibility to be both at once (i.e., a prosumer). This prevents end-users, who are usually consumers, from selling the unused resources they generate.

Our proposed framework address all the weaknesses found in existing utility trading platforms. We address the first major weakness by proposing a multi-utility trading framework where users can exchange different resources using one single framework, currency, and smart contract mechanism using blockchain technology. We also address the second and third weaknesses by using an open architecture where manufacturers can integrate their devices to our proposed framework and entrepreneurs can innovate new business models across supply, pricing, and smart contracts. The common weakness of delegation has already been solved in some of the platforms (5 of the 13). We will adopt the same approach used by these five platforms to address the delegation aspect.

## Proposed Framework

The literature has various proposals for energy exchange platforms [31] [32] [34] [36] [37], water exchange platforms [9] [14], water management systems [7] [8] and many standalone solutions [24] [25] [37]. We cannot neglect the originality of existing solutions, but the business model motivating parties to join in has not been addressed. A consumer is expected to have multiple exchange systems which are often incompatible with each other. This deficiency motivates us to propose the design of a generic utility exchange framework integrating energy, water, Internet services, taxes along with any other needed service. This framework is beneficial from the business model perspective and is the highly effective from a sustainability development point of view as well.

### *Framework Scenarios*

We consider the following scenarios:

**Scenario 1:** in this scenario, an electric car owner purchases a credit of a cryptocurrency issued by the state. The value of this cryptocurrency is stabilized by the issuer which makes it a fiat currency which is suitable for local transactions. This car owner visits a car charging stop managed by a local homeowner, charges his car, and downloads multimedia content while doing the charging. The cost of this service is paid using the fiat currency from his/her electronic wallet on his/her smartphone. The car charging stop is supplied by unused electricity generated from the homeowner's solar panels while the multimedia content is downloaded from the homeowner's home Internet connection. The homeowner uses this credit to pay his/her taxes, Internet, and water bills.

**Scenario 2:** in this scenario, a farmer installs wind turbines on the peripheral of his/her land which generates excess energy not used in the farm. This extra energy is offered on our proposed utility market which is purchased from a nearby factory for a number of fiat currency units. This credit is used by the farmer to cover the water expenses from a nearby irrigator. Similarly, the irrigator uses the received credit to pay his/her taxes and energy used in water irrigation.

**<u>Scenario 3</u>:** in this scenario, a homeowner collects the rainwater from his/her backyard and ceiling and stores it in an aquifer or a rainwater tank [5]. The stored water can be exchanged later for credits to cover the homeowner's taxes. In this case, the homeowner invested once in the rainwater system which in turn helped in covering his/her annual municipal taxes.

Figure 1 depicts the scenarios described above.

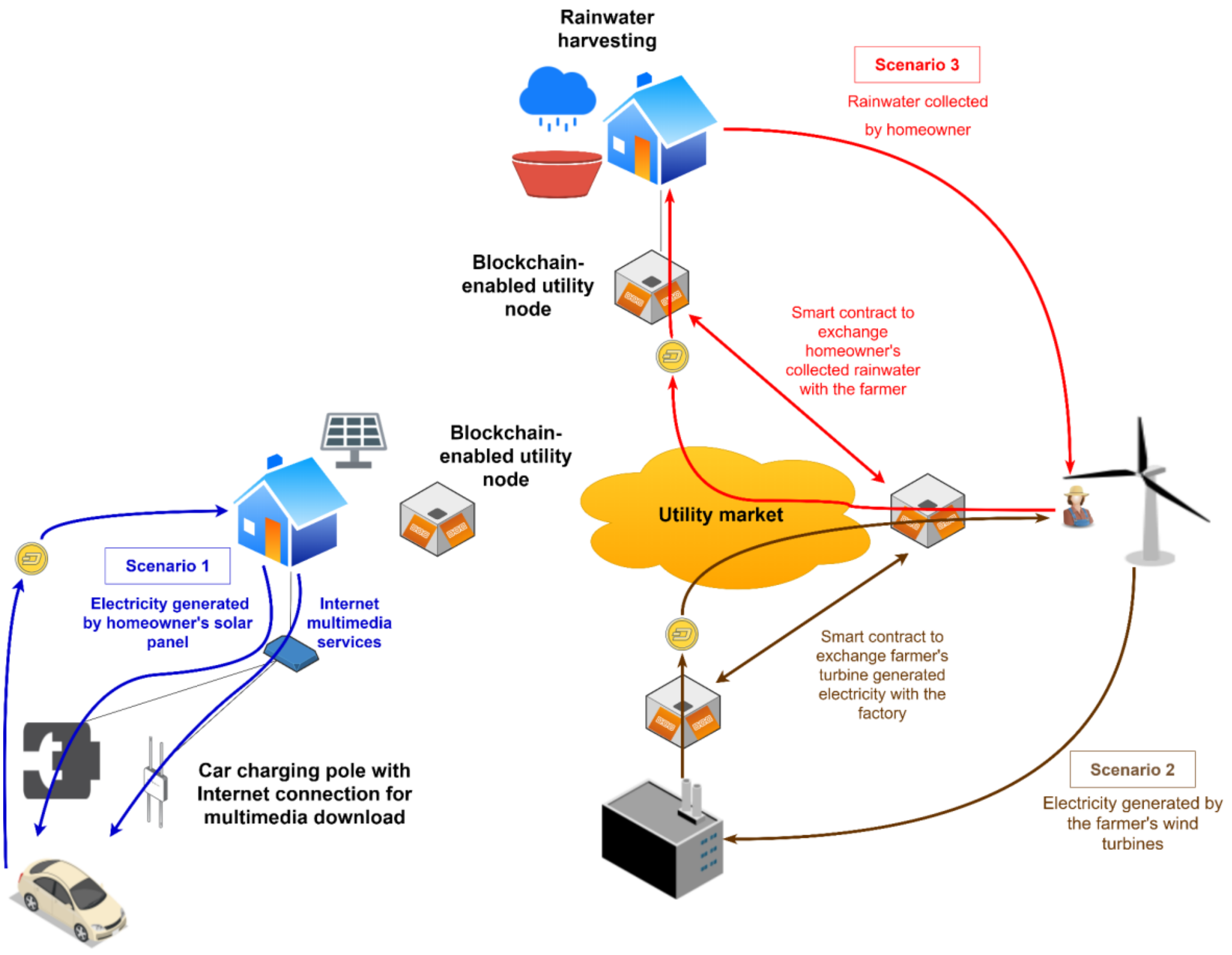

Figure 1. Framework depicting three scenarios

### *Framework Positioning*

This framework allows users to focus on generating their convenient types of resources and replacing them with the types they need, at the locations where they need them and when they need them. A solar panel owner can sell his/her extra energy generated in summer and use the collected fiat currency units to purchase energy in winter from a wind turbine owner. The wind turbine owner can similarly use the collected currency units to purchase energy back from the solar panel owner in the summer.

The proposed framework works as follows:

(1) A consumer installs a utility node connected to our multi-utility market. He/she also installs smart meters on the grid interface, water interface, and on other available services' interfaces. When the user turns his/her washing machine, instead of directly consuming electricity from the electric grid (smart or traditional), the utility node queries the utility market searching for the best price to match its electricity requirements. Once an offer is selected, a smart contract is automatically generated that will transfer the credits once the electricity is exchanged. Similarly, the utility node automatically queries the market searching for a good source of water to cover the requirements of the washing machine. Once an offer is selected another smart contract is generated. When the user's credits are nearly finished, the system send the user a notification, informing him/her to purchase more credits.

(2) A prosumer is a consumer who performs all the actions above but also plays the role of a supplier. The prosumer installs a utility node connected to the blockchain and smart meters to support the available services. He/she also installs a resource generation system, considers a large rainwater tank installed under his/her backyard. His/her utility measures the amount of water stored in the rainwater tank and generates an offer to the market. When the offer is selected, the pump automatically pushes water equal to the required amount and receives the credits after executing the smart contact. When the prosumer comes back home and wants to use electricity, his/her utility node uses the collected credits to cover the cost of the selected offer.

(3) Both the prosumer and the supplier can revoke a contract. The credit settlement is performed based on resources used. The revoking party's reputation decreases after revoking the contract.

The originality of this proposed framework is giving the user the ability to choose his/her convenient resource to generate and replace it with the utilities he/she needs. This motivates users to participate by having to invest in one resource, use one framework, and trade resources with each other using a unified currency system. The proposed framework is a peer-to-peer design, but the state (local government or municipality) can also participate as a user to facilitate and provide any shortage of resources.

A scenario describing the state's intervention has been proposed by Grigoras et al. [7] wherein water suppliers could use Artificial Intelligence (AI) algorithms to offer new water pricing options to reshape demand. In other words, when water supply is available, demand is increased by decreasing price, thus eliminating resource loss. When the supply is not available and water pumping from underground reservoirs is required, prices are increased to push the demand down. Such pricing could achieve significant energy savings and supply the consumers with quality water at an affordable price. Similar dynamic pricing is also applicable to electricity and other resources.

***Framework Requirements***

For this framework to work as expected, several requirements must be satisfied. They include:

*Unique Non-Volatile Currency*

Payment platforms is one of the obstacles facing the widespread deployment of green transportation [3] and volatility is an obstacle facing the widespread of cryptocurrencies. To solve this problem, a stable cryptocurrency is needed to serve as the trade currency in the proposed framework. A fiat currency issued by the state and used to pay the taxes is a suitable solution. A credit-based system is also suitable as proposed by [43] because it also simplifies financial settlements, one of the bottlenecks for peer-to-peer trading [47]. Smart metering and smart contracts complement the role of unique currency in financial settlements. Our proposed framework incorporates a unique non-volatile currency.

*Transparent Pricing*

Transparent pricing of any resource is easily implemented but it is also important for supporting a successful trading platform. If dynamic pricing, transparent pricing, smart metering, and smart contract are implemented, the trading platform can be transparent from a user’s day-to-day operations. Our proposed framework incorporates transparent pricing.

*Smart Contract*

Smart contract is a business logic implemented on and by a blockchain. This business logic includes a set of requirements (conditions) which, if satisfied, will result in the execution of a transaction. In our case, the conditions usually include the proof of utility delivery generated by a trusted smart meter and the utility includes a financial settlement (transfer of funds/credits/currency). Smart contract is a key factor in financial settlement. Our proposed framework incorporates smart contract.

*Smart Metering*

Smart metering is important for source-based reconciliation and smart contract triggering. Smart metering has been frequently used in utility resource trading systems [35] [36] and is considered to be a key factor in financial settlement. Our proposed framework incorporates smart metering.

*Delegated*

Many trading platforms are unidirectional where a user is either a supplier or a consumer without having the possibility of being a prosumer. All existing water trading systems are unidirectional [7] [8] [9] [14] and this limits the opportunities for new business models, user participation in green technologies and bottom-up solutions. Delegated is entrusting the supply possibility to any user and thus transforming all consumers to potential prosumers. Our proposed framework delegates the supply to all users.

*Multi-resource*

One of the problems in trading platforms is their limitation to one resource/utility type. A user is expected to have multiple exchange systems which are often incompatible and this demotivates users from investing and utilizing exchange platforms especially that different systems can use different currencies/credit systems. Our proposed framework enables trading of multiple utilities.

New utility trading systems are not based on blockchain. They often use a centralized approach. Blockchain has been selected as the technology of choice for utility trading systems because of its convenient characteristics which include [3]: decentralization, interconnected autonomy, openness, and intelligence. Our proposed framework uses blockchain.

*Business Model Opportunities for private initiatives*

A flexible and open framework can drive private initiatives to develop new business models to generate profit and further develop the functionalities and services of the considered framework. One of the features that allow private initiatives is the delegation of resource supplies. Other features include the openness in the framework design. Our proposed framework is open to private initiatives and business models.

*Integration between incompatible interfaces*

Device incompatibility is one of the biggest obstacles facing the emergence of green technology [26]. A successful platform should address this issue by proposing an integration middleware to allow proprietary devices to be easily integrated with other systems and interoperate with them. Our proposed framework can provide the required middleware support.

*Security and Privacy*

Security and privacy are also an important obstacle facing the emergence of green technology. The literature contains a number of security and privacy-oriented solutions [32] [36] [37] [46]. Our proposed framework is modular in that it can integrate security and privacy protection measures such as those proposed by [32] [36] [37] [46]. In the future, we plan to add a reputation mechanism that could be used when selecting offers.

***Framework Design***

The proposed framework has 3 main components which are the Utility Node (and Smart meters), the Utility market and the Blockchain miner. The utility node is the device responsible for only one user's utilities. It is connected to the consumer's smart meters which are responsible for measuring and signing the utilities consumed. It is also connected to the producer's smart meters which are responsible for measuring and signing the utilities produced. The utility node includes a digital wallet that stores the cryptocurrency/credit/fiat currency used by the system. The job of the utility node is to:

- Allocate utilities to perform the following tasks:
  - Create offers based on the measurements of the producer’s smart meters.
  - Upload offers to the utility market.
  - Receive order from the utility market and participate in creating the smart contract.
  - Receive the currency after fulfilling the smart contract.
- Satisfy the consumer’s utility needs by executing the following tasks:
  - Meet the user’s utilities’ demand from the utilities he/she produced.
  - Search the utility market for the best offer and participate in creating the smart contract.
  - Read the measurements from the consumer’s smart meter and confirm the fulfilment of the smart contact.
  - Transfer the currency after receiving the utilities.
  - Store service rating in the blockchain related to the smart contract.
- Integrate smart meters and other devices. The node offers an open library that allows drivers to be developed for incompatible devices.

The second main component is the utility market which is a decentralized directory of offers. The offer has a validity period, utility type, utility quantity and price. If the reputation mechanism is implemented, once an offer is received, the utility market searches the blockchain and calculates a reputation index based on the service rating received for the smart contracts where the offer issuer was the supplier. This reputation metric is appended to the offer. Once an offer is selected, the utility market contacts the offer issuer, creates the smart contract, and sends it to the blockchain miners. The supplier's utility node can then ask the blockchain to apply the smart contract when the utility transfer is done. The expired offers are neglected /removed from the utility market.

The third main component is the blockchain (Blockchain Miners). The job of the blockchain is to store smart contracts, apply them, transfer funds, and store service ratings. Figure 2 shows the sequence diagram of the exchanged messages between the framework's main components.

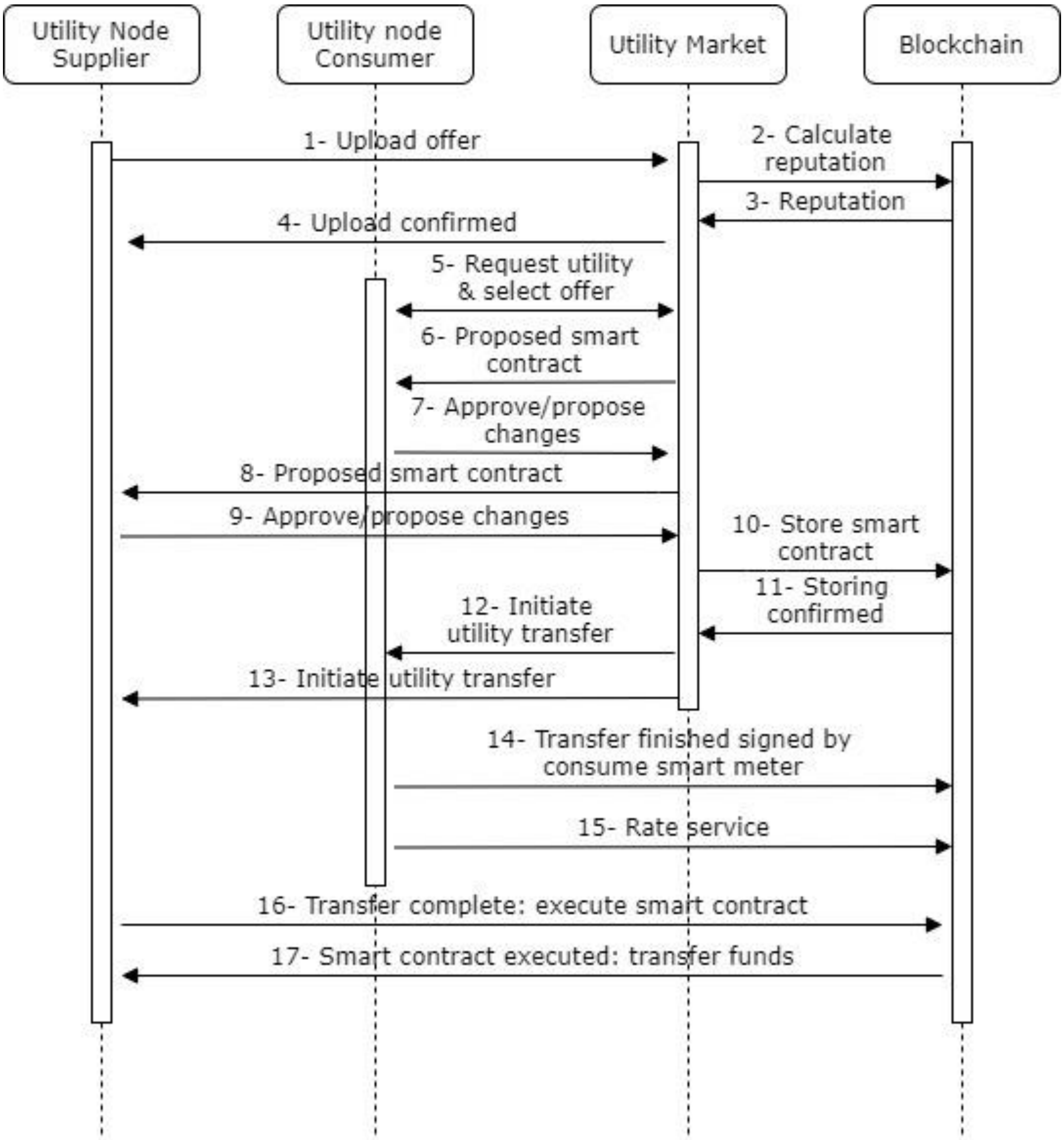

Figure 2. Framework's exchanged messages

Table 2 shows the solution-to-requirements comparison.

Table 2. Requirement fulfilment comparison

| | Fulfilled? | Comment |
|---|---|---|
| *Unique non-volatile currency* | Yes | |
| *Transparent pricing* | Yes | |
| *Smart contract* | Yes | |
| *Smart metering* | Yes | |
| *Delegated* | Yes | |
| *Multi-resource* | Yes | |
| *Blockchain-based* | Yes | |
| *Business model opportunities for private initiatives* | Yes | Smart contract negotiation, messages 6,7, 8, and 9 |
| *Integration between incompatible interfaces* | Yes | Utility node offers open library for integration |
| *Security, privacy and reputation* | Yes | Messages 2, 3, and 15. The proposed framework can integrate the security solutions proposed by [32] [36] [37] [46] |

## Framework Implementation Challenges

The proposed framework has been intentionally designed with some abstractions to provide the implementer with the flexibility needed to address some of the challenges faced when implementing the framework. These challenges include:

- **Selecting the blockchain Type**: the implemented blockchain type considerably changes the dynamicity of the supported smart contracts and thus how innovative can private initiatives be. Using a blockchain with extensive support to customized smart contracts allows new business models to thrive thereby attracting interests in the framework. The implemented blockchain also affects the exchange currency used.
- **Selecting credit-system or cryptocurrency-system**: theoretically, both systems work, but the complexity of implementing each system depends on the blockchain type used. Additionally, each system has its impact on the “Business Model Opportunities for private initiatives” and consequently the adoption of the framework. It is important to consider the penetration rate of such systems especially when the people in developing countries usually need a motivation other than how environment-friendly the solution is to break the barriers to adoption [44].

- **Deciding on credit loans**: integrating the framework with peer-to-peer lending services or banks can be an interesting idea [40] especially when the banks can drive the framework's implementation and adoption. Banks can also fund private startups developing services for this framework. Although this feature looks interesting and aligned with the framework's goal in not relying on state's maintenance and management of the framework, but it may introduce a challenge to normal users. Wealthy investors, such as banks, have the capability of generating cheaper resources (due to economy of scale) that would defer normal users from generating and selling their unused resources.

## Conclusion

In this paper, we have proposed the first multi-utility trading framework which can serve the needs of developing countries. Typical solutions for developing countries should require minimal investment and maintenance from the state and thrive on private initiatives. We proposed an adoptable framework by considering all the obstacles faced by existing trading platforms. These obstacles include: integration with incompatible devices, opportunities to innovate new business models, and multiple platforms for different utilities.

In addition to the focus on the proposed framework's adoptability, its originality includes satisfying the 10 requirements needed by a successful trading system, which are not satisfied by any other platform to date. Our proposed framework's design also includes some abstractions that provide flexibility in addressing existing challenges and future requirements.

Future extension of the proposed framework could include the addition of a reputation mechanism that could be used when selecting offers. This reputation mechanism benefits from blockchain's immutability which is swell suited for reputation and rating systems. We plan to consider applying differential privacy to protect the users' identity as part of our future works.